# Percentile rank scores are congruous indicators of relative performance, or aren't they?

Ronald Rousseau

KHBO (Association K.U.Leuven), Faculty of Engineering Technology,
Zeedijk 101, B-8400 Oostende, Belgium
E-mail: ronald.rousseau@khbo.be

K.U.Leuven, Dept. Mathematics,
Celestijnenlaan 200B, B-3000 Leuven (Heverlee), Belgium

Universiteit Antwerpen, IBW, Venusstraat 35,
B-2000 Antwerpen, Belgium

Abstract

Percentile ranks and the I3 indicator were introduced by Bornmann, Leydesdorff, Mutz and Opthof. These two notions are based on the concept of percentiles (or quantiles) for discrete data. As several definitions for these notions exist we propose one that we think is suitable in this context. Next we show that if the notion of relative congruous indicators is carefully defined then percentile rank scores are congruous indicators of relative performance. The I3 indicator is a strictly congruous indicator of absolute performance.

**Introduction**

During recent years the field of scientometrics and especially that part involved in research evaluation has gone through a number of major shocks when old truths and methods do not seem to be valid anymore (Lundberg, 2007; Opthof & Leydesdorff, 2010; van Raan et al., 2010). How should indicators for research evaluation be calculated? Can arithmetic averages still play a prominent role? Because of problems with the so-called crown indicator and even with the standard (and other synchronous) impact factors (Rousseau & Leydesdorff, 2011) colleagues have begun a search to find other indicators that do not depend on averages.

**An axiomatic approach**

Following the lead of mathematicians, statisticians, econometricians and ecologists (Dalton, 1920; Hardy, Littlewood & Pólya, 1952; Pielou, 1975; Stirling, 2007) also informetricians have taken recourse to axiomatic approaches. Examples of such approaches are the study of the informetric laws (Egghe, 2005), inequality (Egghe, 2005; Rousseau, 1992), ranking journals (Palacios-Huerta & Volij, 2004) and h-type indices (Marchant, 2009; Quesada, 2009; Rousseau, 2008; Woeginger, 2008). When it seemed that there might be problems with the original crown indicator it was an axiomatic approach that dealt the final blow. Indeed, one of the requirements for an acceptable indicator is that it be consistent (for the definition of this notion we refer the reader to (Waltman *et al.*, 2011a)). However, in (Waltman et al., 2011a) it is shown that the (old) crown indicator (a ratio of averages) does not satisfy this requirement, while the Karolinska indicator (an average of ratios) does. Yet, also averages of ratios may present some practical problems as illustrated in (Waltman et al., 2011b). The property we will use is based on (Waltman & Van Eck, 2009a; Waltman et al., 2011a). However, contrary to their definition, the notion of expected number of citations does not play a role at all in our approach. It is also very close to the notion of independence as defined by Bouyssou & Marchant (2011). The exact difference between consistency and independence is discussed in (Bouyssou & Marchant, 2011).

As we will slightly change the definition of independence (and consistency) we refer to it by a different name, using *congruous* instead of *consistent*.

Definition 1. Congruous indicators of average performance

Let A and B denote sets of documents, containing the same number of documents (#A = #B). Let A' and B' be the sets A and B to which the same document has been added. Then an indicator of average performance f is said to be strictly congruous if:

$$f(A) > f(B) \text{ if and only if } f(A') > f(B') \qquad (1)$$

It is said to be congruous if:

$$f(A) > f(B) \Rightarrow f(A') \geq f(B') \qquad (2)$$

Clearly, if an indicator of average performance is strictly congruous it is also congruous. The opposite is not true. Consider, for example, the indicator F(A) = Max(3, average number of citations of articles in A). Then, if F(A) > F(B) it is easy to add to both sets an article with a large number of citations leading to F(A') = F(B') = 3.

Definition 2. Congruous indicators of total performance

Let A and B denote sets of documents and let A' and B' be the sets A and B to which the same document has been added. Then an indicator of total performance f is said to be strictly congruous if:

$$f(A) > f(B) \text{ if and only if } f(A') > f(B') \qquad (3)$$

**Percentile rank scores**

Consider a set A and a reference set S containing all elements in A, hence A ⊆ S. Moreover, we assume that a function g from S to the positive real numbers is given, leading to the multiset g(S). Note that we consider g(S) as a multiset, as we consider the images g(s), s in S, as separate entities (even if their values are the same). A standard situation is the case that A consists of a set of articles, the set S consists of all articles in the journals in which the set A is published (published in the same year), and a function g which maps an article to the number of citations it has received over a given period (and there may be several articles with the same number of citations).

Now a rule is given which subdivides the set S into K disjoint classes, based on the values of the function g. If a document belongs to class k then it receives a score $x_k$. Note that this score only depends on the class (and hence on S), but not on A. Again a standard situation is the case that there are 100 percentile classes (or ten decile classes). In the case of percentiles articles belonging to the top 1% receive a score of 100, those belonging to the top 2% (and not to the top 1%) receive a score of 99, and so on.

Definition 3. Percentile rank scores (Bornmann & Mutz, 2011; Leydesdorff et al., 2011)

Let A be a set of N documents and let $n_A(k)$ (or simply n(k) when the set A is of no importance) be the number of documents in A that belong to class k.

Then the percentile rank score of A is defined as:

$$R(A) = \sum_{k=1}^{K} x_k \frac{n_A(k)}{N} \qquad (4)$$

Clearly, the value of R(A) depends not only on A, but also on the reference set S, the K classes used and their score.

**A note on percentiles and related classes**

Percentile rank scores use percentiles and, usually, classes based on percentiles. Hence these two notions must be clarified. There is, however, no general agreement about how to define percentiles for discrete data, see e.g. (Hyndman & Fan, 1996). We follow the approach by (Beirlant et al., 2005). This definition is simple and has the advantage to always give the highest score to the best performer in a reference set. We provide an example illustrating the used concepts and compare with the definition as used in (Leydesdorff et al., 2011). Yet, in the appendix, we will also use another definition (Egghe & Rousseau, 2001) and illustrate the difference.

Given a finite multiset of numbers (some numbers may occur several times, hence this is a multiset and not necessarily a set), *M = [$x_i$, i = 1,..,n]*, ranked from smallest to largest, we define for $x \in \mathbb{R}$:

$$\widehat{F_n}(x) = \frac{1}{n} \#\{x_i \in N;\ x_i \leq x; i = 1, \dots, n\}$$

This function is a right-continuous step function with discontinuities in the points $x_i$. If M = [1,15,79,2,3,11,15,15,185, 47, 71, 18, 101] or in ranked order: M = [ $x_1$ = 1, $x_2$ =

2, $x_3 = 3$, $x_4 = 11$, $x_5 = 15$, $x_6 = 15$, $x_7 = 15$, $x_8 = 18$, $x_9 = 47$, $x_{10} = 71$, $x_{11} = 79$, $x_{12} = 101$, $x_{13} = 185$]; n = 13. Then, for the example multiset M the values for the corresponding function $\widehat{F_{13}}(x)$ are shown in Table 1.

Table 1. $\widehat{F_n}(x)$ - values

$$\widehat{F_{13}}(x) = 0, if\ x < 1;$$

$$\widehat{F_{13}}(x) = \frac{1}{13} \approx 0.077, if\ 1 \leq x < 2; ;$$

$$\widehat{F_{13}}(x) = \frac{2}{13} \approx 0.154,\ if\ 2 \leq x < 3; ;$$

$$\widehat{F_{13}}(x) = \frac{3}{13} \approx 0.231, if\ 3 \leq x < 11;$$

$$\widehat{F_{13}}(x) = \frac{4}{13} \approx 0.308, if\ 11 \leq x < 15;$$

$$\widehat{F_{13}}(x) = \frac{7}{13} \approx 0.538, if\ 15 \leq x < 18;$$

$$\widehat{F_{13}}(x) = \frac{8}{13} \approx 0.615, if\ 18 \leq x < 47;$$

$$\widehat{F_{13}}(x) = \frac{9}{13} \approx 0.692, if\ 47 \leq x < 71;$$

$$\widehat{F_{13}}(x) = \frac{10}{13} \approx 0.769, if\ 71 \leq x < 79;$$

$$\widehat{F_{13}}(x) = \frac{11}{13} \approx 0.846, if\ 79 \leq x < 101;$$

$$\widehat{F_{13}}(x) = \frac{12}{13} \approx 0.923, if\ 101 \leq x < 185;$$

$$\widehat{F_{13}}(x) = 1, if\ x \geq 185.$$

Let now 0 < p ≤ 1, then the p[th]-quantile, denoted as $\widehat{Q_n}(p)$ is a kind of inverse of $\widehat{F_n}(x)$. However, as $\widehat{F_n}(x)$ is not injective an inverse in the strict sense cannot be defined. Consequently, a choice must be made and following (Beirlant et al., 2005) we define

$$\widehat{Q_n}(p) = min\ \{x\ ;\ \widehat{F_n}(x) \geq p\}$$

For p = 0 we put $\widehat{Q_n}(0) = min\{x_i\} = x_1$. When p takes the values 0, 0.01, 0.02, …, 0.99, 1.00 then these quantiles are called percentiles. We note that, in particular $\widehat{Q_n}(1) = max\{x_i\} = x_n$. Also the function $\widehat{Q_n}(p)$ is a step function with discontinuities in the points $x_i$; in these points the function is left-continuous.

For our example $\widehat{Q_{13}}(p) = 1$ for $p = 0, 0.01, …, 0.07$ ; $\widehat{Q_{13}}(p) = 2$ for $p = 0.08, …, 0.15$ ; $\widehat{Q_{13}}(p) = 3$ for $p = 0.16, …, 0.2$; $\widehat{Q_{13}}(p) = 11$ for $p = 0.24, …, 0.30$ ; $\widehat{Q_{13}}(p) = 15$ for $p = 0.31, …, 0.53$ ; $\widehat{Q_{13}}(p) = 18$ for $p = 0.54, …, 0.61$ ; $\widehat{Q_{13}}(p) = 47$ for $p = 0.62, …, 0.69$ ;

$\widehat{Q_{13}}(p) = 71$ for $p = 0.70, ...,0.76$ ; $\widehat{Q_{13}}(p) = 79$ for $p = 0.77, ...,0.84$ ; $\widehat{Q_{13}}(p) = 101$ for $p = 0.85, ...,0.92$ ; $\widehat{Q_{13}}(p) = 185$ for $p = 0.93, ...,1.00$ .

Percentile values can be used to delineate classes. Examples may be: using hundred or ten classes of equal breadth, or six classes, such as the NSF categories, of unequal breadth.

We propose using the following percentile classes and scores in g(S): the 99th percentile class is $[\widehat{Q_n}(0.99), \widehat{Q_n}(1.0)]$ with score 100; the 98th percentile class is $[\widehat{Q_n}(0.98), \widehat{Q_n}(0.99)[$ with score 99; and generally the tth percentile class (t = 0, 1,..., 99) is $[\widehat{Q_n}\left(\frac{t}{100}\right), \widehat{Q_n}\left(\frac{t+1}{100}\right)[$ with score t+1. The last percentile class (t=0) is class $[\widehat{Q_n}(0), \widehat{Q_n}(0.01)[$ with score 1. Note that these intervals are right-open, except for the 99th percentile class which is a closed interval. In this way the largest number in the set always belongs to the highest percentile. We further note that for a small number of discrete data most of these intervals will be empty. Similarly, decile classes are defined as $[\widehat{Q_n}(0.9), \widehat{Q_n}(1.0)]$ with score 10; $[\widehat{Q_n}(0.8), \widehat{Q_n}(0.9)[$ with score 9 and the k-th class (k = 0, 1, 2,..., 9) $[\widehat{Q_n}\left(\frac{k}{10}\right), \widehat{Q_n}\left(\frac{k+1}{10}\right)[$ with score k+1.

In our example the percentile scores are: $x_{13}$ = 185 has score 100, $x_{12}$ = 101 has score 93; $x_{11}$ = 79 has score 85; $x_{10}$ = 71 has score 77; $x_9$ = 47 has score 70; $x_8$ = 18 has score 62; $x_5 = x_6 = x_7$ = 15 have score 54; $x_4$ = 11 has score 31; $x_3$ = 3 has score 24; $x_2$ = 2 has score 16 and $x_1$ = 1 has score 8.

Indeed, let us calculate, as an example $[\widehat{Q_n}(0.92), \widehat{Q_n}(0.93)[$. This interval is equal to $[min\{x\,;\,\widehat{F_n}(x) \geq 0.92\}, min\{x\,;\,\widehat{F_n}(x) \geq 0.93\}[$ = $[101,185[$ (reading values from Table 1). Hence 101 belongs to the 92nd percentile class, and hence its percentile score is 93. We further note that for our example the 93rd percentile class is empty. Indeed $[\widehat{Q_n}(0.93), \widehat{Q_n}(0.94)[$ = $[min\{x\,;\,\widehat{F_n}(x) \geq 0.93\}, min\{x\,;\,\widehat{F_n}(x) \geq 0.94\}[$ = $[185,185[$ = $\emptyset$.

How do Leydesdorff et al. (2011) determine percentiles? The quantile of paper $x_j$ is defined as $\#\{x_m: x_m < x_j\} * \frac{100}{n}$. Paper $x_j$ then belongs to percentile class $\left\lfloor \#\{x_m: x_m < x_j\} * \frac{100}{n} \right\rfloor$ (where the symbol $\lfloor z \rfloor$ denotes the largest integer smaller than or equal to z) with score $\left\lfloor \#\{x_m: x_m < x_j\} * \frac{100}{n} \right\rfloor$ +1. As an example of this procedure we determine the percentile score of $x_8$ = 18. According to the *Leydesdorff et al.* procedure we first determine $\#\{x_m: x_m < x_j\} * \frac{100}{n}$ = 7*100/13 = 53.846. This implies that $x_8$ belongs to percentile class 53 and hence has a score equal to 54. Note that in our approach above the score is 62. As the Leydesdorff et al. procedure shifts all scores this leads to the problem that the best performer does not have a score of 100 (or any other chosen highest score), but that it heavily depends on n. In our approach no ad hoc procedure (e.g. adding 0.9) is necessary to compensate for this dependence, as in (Leydesdorff & Bornmann, 2011).

Note that the following alternative procedure easily leads to the same values as in the more formal approach explained above. Determine $\#\{x_m: x_m \leq x_j\} * \frac{100}{n}$. (the only difference with the *Leydesdorff et al.* procedure is that we added an equality sign). Paper $x_j$ then belongs to percentile class $\lfloor \#\{x_m: x_m \leq x_j\} * \frac{100}{n} \rfloor$ with score $\lfloor \#\{x_m: x_m \leq x_j\} * \frac{100}{n} \rfloor$ +1 (for $x_n$ there is no increase by one). As an example of this procedure we again determine the percentile score of $x_8$ = 18. According to this alternative procedure we first determine $\#\{x_m: x_m \leq x_j\} * \frac{100}{n}$ = 8*100/13 = 61.54. This implies that $x_8$ belongs to percentile class 61 and hence has a score equal to 62.

**Percentile rank scores are strictly congruous indicators of average performance**

The truth of this proposition does not depend on the ways percentiles and percentile classes are calculated, as long as these are fixed. Yet, there is a finer point to be observed which is made further on.

Proposition. Percentile rank scores are strictly congruous indicators of average performance

Proof. Assume that $R(A) = \sum_{k=1}^{K} x_k \frac{n_A(k)}{N} > R(B) = \sum_{k=1}^{K} x_k \frac{n_B(k)}{N}$ (recall that A and B must have the same number of elements). We now add to A and to B a document that belongs to class j and denote the new sets by A' and B'. Then $R(A') = \frac{N}{N+1} R(A) + x_j \frac{1}{N+1}$, while $R(B') = \frac{N}{N+1} R(B) + x_j \frac{1}{N+1}$. Removing a common document from the sets A and B leads to a similar conclusion. This shows that R is a strictly congruous indicator of average performance.

Percentile rank scores are not congruous indicators of absolute performance. Note though that they are not meant to be!

Example

Let A consist of two articles: one with score 1 and one with score 4. Its percentile rank score is (1+4)/2 = 2.5. Let B consist of four articles: two with score 1, one with score 3 and one with score 6. Its percentile rank score is (1+1+3+6)/4 = 2.75

Now we add to A and B the same article with score 5.

The new percentile rank score of A is now: (1+4+5)/3 = 3.33

The new percentile rank score of B is: (1+1+3+5+6)/5 = 16/5 = 3.2

Although we have shown that percentile rank scores are strictly congruous indicators of average performance, there is, however, a way to violate the requirement of congruousness for relative indicators. This is the finer point we mentioned above. Let us consider the following counterexample.

**A counterexample**

Assume that classes are determined as deciles. Those in the highest decile receive a score of 10; the next 10% a score of 9 and so on.

The reference set consists of 99 articles which have received 99, 98 ... 1 citation(s). The function $\widehat{F_{99}}(x) = \frac{\lfloor x \rfloor}{99}$, for $0 \leq x \leq 99$. The 9$^{th}$ decile is then $[\widehat{Q_{99}}(0.9), \widehat{Q_{99}}(1.0)] = [90,99]$, with score 10; the 8$^{th}$ decile is $[\widehat{Q_{99}}(0.8), \widehat{Q_{99}}(0.9)[ = [80, 90[$, with score 9 and so on.

Group A consists of the articles receiving 96,86,76,66,56,46,36,26,16,6 citations. Group A has a percentile rank score of 55/10 = 5.5 .

Group B consists of the articles receiving 89, 88, 79, 69, 59, 49, 39, 29, 19, 9 citations. Group B has a percentile rank score of 54/10 = 5.4, so that group A has a higher score than group B.

Now we add a new article to A, B and to the reference set. This article has no citations. For the new situation we have: $\widehat{F_{100}}(x) = \frac{\lfloor x \rfloor + 1}{100}$, for $0 \leq x \leq 99$. The 9$^{th}$ decile is then $[\widehat{Q_{100}}(0.9), \widehat{Q_{100}}(1.0)] = [89,99]$, with score 10; the 8$^{th}$ is $[\widehat{Q_{100}}(0.8), \widehat{Q_{100}}(0.9)[ = [79, 89[$, with score 9 and so on.

The new score of A is (55+1)/11 ≈ 5.09. However, the new score for B is: (10+9+9+8+...+2+1)/11 = 64/11 ≈ 5.82.

Hence, by adding an article without any citations, B's percentile rank score became higher than A's and even higher than A's original score.

We admit that this counterexample is of a purely theoretical nature. The problem that occurred can easily be avoided by stipulating in the definition of congruousness that the added article d must already belong to the intersection of the reference sets for A and B, hence adding d does not change any of the two reference sets, and hence also classes derived from percentiles do not change. Consequently, we propose to add this requirement to the definition of congruousness. This leads to the following definition.

Definition 1a. Congruous indicators of average performance

Let A and B denote sets of documents, containing the same number of documents (#A = #B). Let A $\subset$ S (the reference set of A) and let B $\subset$ T (the reference set of B). Let A' and B' be the sets A and B to which the same document $d \in S \cap T$ has been added. Then an indicator of average performance f is said to be strictly congruous if:

$$f(A) > f(B) \text{ if and only if } f(A') > f(B') \tag{5}$$

**The I3 indicator**

The I3 indicator (Leydesdorff & Bornmann, 2011), where I3 stands for Integrated Impact Indicator, is defined in a similar way as the percentile rank score as given in equation (4). No division by N is performed, but the role of the reference set S is the same. Hence, using the notation introduced above, we have the following definition.

Definition 4. The I3 score of a set A is defined as:

$$I3(A) = \sum_{k=1}^{K} x_k n_A(k) \tag{6}$$

Proposition. I3 scores are strictly congruous indicators of absolute performance

Proof. Assume that $I3(A) = \sum_{k=1}^{K} x_k n_A(k) > I3(B) = \sum_{k=1}^{K} x_k n_B(k)$ (now A and B do not necessarily have the same number of elements). Adding to sets A and B a document that belongs to class j and denoting the new sets by A' and B' gives: $I3(A') = I3(A) + x_j$, while $I3(B') = I3(B) + x_j$. This trivially shows that I3 is a strictly congruous indicator of absolute performance.

**Notes**

Note 1

As observed by Leydesdorff and Bornmann the I3 score of a set A can also be defined without the notion of classes. In that case one just uses the quantile value of each element in A (with respect to the reference set S). This does not change the fact that the I3 indicator is a strictly congruous indicator of absolute performance.

Note 2

The proposal by Waltman and van Eck (2009b), see also (Plomp, 1990) to use the number of highly cited publications as an indicator (the HCP-indicator) can be consider as a special case of the I3 score. Indeed, taking only two classes (the highly cited ones and the other ones) and giving the highly-cited ones a score of 1, and the other ones a score of 0, reduces formula (6) to the number of highly-cited publications in a set A.

**Conclusion**

When the notion of relative congruous indicators is carefully defined then percentile rank scores are congruous indicators of relative performance. Similarly, the I3 indicator is a strictly congruous indicator of absolute performance. The HCP-indicator can be considered as a special (simple) case of the I3 indicator.

Acknowledgements. The author thanks Loet Leydesdorff for many useful discussions about the notions studied in this article. He also thanks Yuxian Liu, Leo Egghe, Raf Guns and Ying (Fred) Ye for helpful suggestions. Work of the author is supported by NSFC grants 70773101 and 7101017006.

Appendix

As said before, there exist many definitions for the notion of a quantile. In this appendix we compare the definition as proposed e.g. in (Beirlant et al., 2005) with the one used, e.g., in (Egghe & Rousseau, 2001).

For $0 \leq p \leq 1$, quantiles in the sense of Egghe & Rousseau (2001), denoted here as $\widehat{P}_n(p)$, can be defined as $\widehat{P}_n(p) = \frac{\lim_{y<p} \widehat{Q_n(y)} + \lim_{y>p} \widehat{Q_n(y)}}{2}$ ($0 < p < 1$) with $\widehat{P}_n(0) = \widehat{Q}_n(0)$ and $\widehat{P}_n(1) = \widehat{Q}_n(1)$. Limits of a function in a point p are calculated by first removing this point p from the domain of definition and then calculating the limit (in this case the left-hand or the right-hand limit)[1]. If $\widehat{Q}_n(x)$ is continuous in the point p then $\widehat{Q}_n(p) = \widehat{P}_n(p)$. Hence, there can only be a difference between these two notions in points where the quantile function $\widehat{Q}_n$ is discontinuous, this means in points x of the form x = m/n, 0 < m < n, with m a natural number. In the example used in this article, n = 13, hence m/n is never a multiple of 1/100. So for the example we have used percentile values coincide for the two definitions. Next we present an example where percentile values (and deciles) do not coincide. We take n = 10.

Let $S_0$ be a set of articles with corresponding citations [1,3,7,8,8,12,17,23,30,60]. Then Table 2 gives the values of $\widehat{F}_{10}(x)$.

Table 2

|  | $\widehat{F}_{10}(x)$ |  | $\widehat{F}_{10}(x)$ |
|---|---|---|---|
| x < 1 | 0 | 12 ≤ x < 17 | 0.6 |
| 1 ≤ x < 3 | 0.1 | 17 ≤ x < 23 | 0.7 |
| 3 ≤ x < 7 | 0.2 | 23 ≤ x < 30 | 0.8 |
| 7 ≤ x < 8 | 0.3 | 30 ≤ x < 60 | 0.9 |
| 8 ≤ x < 12 | 0.5 | 60 ≤ x | 1 |

Now $\widehat{Q}_{10}$ is defined as:

$\widehat{Q}_{10}(p) = 1$ for $0 \leq p \leq 0.10$

$\widehat{Q}_{10}(p) = 3$ for $0.10 < p \leq 0.20$

$\widehat{Q}_{10}(p) = 7$ for $0.20 < p \leq 0.30$

$\widehat{Q}_{10}(p) = 8$ for $0.30 < p \leq 0.50$

$\widehat{Q}_{10}(p) = 12$ for $0.50 < p \leq 0.60$

---

[1] There exist different definitions of limits in a point. However, to the best of our knowledge these definitions coincide when the point p is first removed. This is the reason for this special procedure.

$\widehat{Q_{10}}(p) = 17$ for $0.60 < p \leq 0.70$

$\widehat{Q_{10}}(p) = 23$ for $0.70 < p \leq 0.80$

$\widehat{Q_{10}}(p) = 30$ for $0.80 < p \leq 0.90$

$\widehat{Q_{10}}(p) = 60$ for $0.90 < p \leq 1.00$

while $\widehat{P_{10}}$ is defined as:

$\widehat{P_{10}}(p) = 1$ for $0 \leq p < 0.10$; $\widehat{P_{10}}(p) = 2$ for $p = 0.10$;

$\widehat{P_{10}}(p) = 3$ for $0.10 < p < 0.20$; $\widehat{P_{10}}(p) = 5$ for $p = 0.20$;

$\widehat{P_{10}}(p) = 7$ for $0.20 < p < 0.30$; $\widehat{P_{10}}(p) = 7.5$ for $p = 0.30$;

$\widehat{P_{10}}(p) = 8$ for $0.30 < p < 0.50$; $\widehat{P_{10}}(p) = 10$ for $p = 0.50$;

$\widehat{P_{10}}(p) = 12$ for $0.50 < p < 0.60$; $\widehat{P_{10}}(p) = 14.5$ for $p = 0.60$;

$\widehat{P_{10}}(p) = 17$ for $0.60 < p < 0.70$; $\widehat{P_{10}}(p) = 20$ for $p = 0.70$;

$\widehat{P_{10}}(p) = 23$ for $0.70 < p < 0.80$; $\widehat{P_{10}}(p) = 26.5$ for $p = 0.80$;

$\widehat{P_{10}}(p) = 30$ for $0.80 < p < 0.90$; $\widehat{P_{10}}(p) = 45$ for $p = 0.90$;

$\widehat{P_{10}}(p) = 60$ for $0.90 < p < 1.00$

Corresponding decile classes are:

$[\widehat{Q_{10}}(0.9), \widehat{Q_{10}}(1.0)] = [30, 60] = \{30,60\}$ with score 10

$[\widehat{Q_{10}}(0.8), \widehat{Q_{10}}(0.9)[ = [23, 30[ = \{23\}$ with score 9

$[\widehat{Q_{10}}(0.7), \widehat{Q_{10}}(0.8)[ = [17, 23[ = \{17\}$ with score 8

$[\widehat{Q_{10}}(0.6), \widehat{Q_{10}}(0.7)[ = [12, 17[ = \{12\}$ with score 7

$[\widehat{Q_{10}}(0.5), \widehat{Q_{10}}(0.6)[ = [8, 12[ = \{8,8\}$ (this must be considered as a multiset) with score 6

$[\widehat{Q_{10}}(0.4), \widehat{Q_{10}}(0.5)[ = \emptyset$

$[\widehat{Q_{10}}(0.3), \widehat{Q_{10}}(0.4)[ = [7, 8[ = \{7\}$ with score 4

$[\widehat{Q_{10}}(0.2), \widehat{Q_{10}}(0.3)[ = [3, 7[ = \{3\}$ with score 3

$[\widehat{Q_{10}}(0.1), \widehat{Q_{10}}(0.2)[ = [1, 3[ = \{1\}$ with score 2

$[\widehat{Q_{10}}(0), \widehat{Q_{10}}(0.1)[ = \emptyset$

Based on $\widehat{P_{10}}$ we have the following deciles and decile scores:

$[\widehat{P_{10}}(0.9), \widehat{P_{10}}(1.0)]$ = [45 , 60] = {60} with score 10

$[\widehat{P_{10}}(0.8), \widehat{P_{10}}(0.9)[$ = [26.5 , 45 [ = {30} with score 9

$[\widehat{P_{10}}(0.7), \widehat{P_{10}}(0.8)[$ = [20 , 26.5 [ = {23} with score 8

$[\widehat{P_{10}}(0.6), \widehat{P_{10}}(0.7)[$ = [14.5 , 20 [ = {17} with score 7

$[\widehat{P_{10}}(0.5), \widehat{P_{10}}(0.6)[$ = [10 , 14.5 [ = {12} with score 6

$[\widehat{P_{10}}(0.4), \widehat{P_{10}}(0.5)[$ = [8 , 10 [ = {8,8} (again a multiset) with score 5

$[\widehat{P_{10}}(0.3), \widehat{P_{10}}(0.4)[$ = [7.5 , 8 [ = Ø

$[\widehat{P_{10}}(0.2), \widehat{P_{10}}(0.3)[$ = [5 , 7.5 [ = {7} with score 3

$[\widehat{P_{10}}(0.1), \widehat{P_{10}}(0.2)[$ = [2 , 5 [ = {3} with score 2

$[\widehat{P_{10}}(0), \widehat{P_{10}}(0.1)[$ = [0, 2 [ = {1} with score 1

This example illustrates the difference between these two definitions of percentiles.

The counterexample we presented is not valid anymore when using the function $\widehat{P_n}(p)$. Hence we provide a slight variation.

Consider the set *g(T)* = {0,1,…,99} as the set of citations (n = 100) received by a reference set *T*. Then $\widehat{F_{100}}(x) = \frac{\lfloor x \rfloor + 1}{100}$, for 0 ≤ x ≤ 99, while $\widehat{F_{100}}(x) = 0$ for $x < 0$ and $\widehat{F_{100}}(x) = 1$ for $x > 99$. Consequently $\widehat{Q_{100}}(p) = (100p) - 1$ for *p = 0.01, 0.02, …, 0.99, 1*, while $\widehat{P_{100}}(p) = \frac{(100p-1)+(100(p+0.01)-1)}{2} = (100p) - 0.5$, for *p = 0.01, 0.02, …, 0.99*; and $\widehat{P_{100}}(1.00) = 99$. Then $[\widehat{Q_{100}}(0.9), \widehat{Q_{100}}(1.0)]$ = [89, 99], with score *10*; and similarly $[\widehat{Q_{100}}(d), \widehat{Q_{100}}(d+0.1)[$ = *[100d-1, 100(d+0.1)-1[*, for *d = 0.1, …, 0.8* with score *10(d+0.1)* and $[\widehat{Q_{100}}(0), \widehat{Q_{100}}(0.1)[$ = *[0, 9[*, with score *1*. Further: $[\widehat{P_{100}}(0.9), \widehat{P_{100}}(1.0)]$ = *[89.5, 99]* with score *10* and similarly $[\widehat{P_{100}}(d), \widehat{P_{100}}(d+0.1)[$ = *[100d-0.5, 100(d+0.1)-0.5[, for d = 0.1, …, 0.8* with score *10(d+0.1)* and $[\widehat{P_{100}}(0), \widehat{P_{100}}(0.1)[$ = *[0, 9.5[* with score *1*.

When group A consists of the articles receiving 96,86,76,66,56,46,36,26,16,6 citations then its percentile rank score is *55/10 = 5.5* ; while group B, consisting of the articles receiving *89, 88, 79, 69, 59, 49, 39, 29, 19, 9* citations has a percentile rank score of *54/10 = 5.4*, so that group A has a higher score than group B.

Now adding an article with zero citations leads to the new reference set $T_0$ consisting of articles with citations *[0,0,1,…,99]*. Then $\widehat{F_{101}}(x) = \frac{\lfloor x \rfloor + 2}{101}$, for *0 ≤ x ≤ 99*. Consequently $\widehat{Q_{101}}(p) = \widehat{P_{101}}(p)$ (see note below) $= \lceil 101 * p - 2 \rceil = 100 * p - 1$ for *p*

= *0.01, 0.02, …, 0.99*, where $\lceil t \rceil$ denotes the smallest integer larger than or equal to *t*. Moreover, we put $\widehat{P_{101}}(0) = \widehat{Q_{101}}(0) = 0$ and $\widehat{P_{101}}(1) = \widehat{Q_{101}}(1) = 99$. Then $[\widehat{Q_{101}}(0.9), \widehat{Q_{101}}(1.0)] = [\widehat{P_{101}}(0.9), \widehat{P_{101}}(1.0)]$ = *[89, 99]*, with score *10*; and similarly $[\widehat{Q_{101}}(d), \widehat{Q_{101}}(d+0.1)[ = [\widehat{P_{101}}(d), \widehat{P_{101}}(d+0.1)[$ = *[100d-1, 100(d+0.1)-1[*, for *d* = *0.1, …, 0.8* with score *10(d+0.1)* and finally $[\widehat{Q_{101}}(0), \widehat{Q_{101}}(0.1)[ = [\widehat{P_{101}}(0), \widehat{P_{101}}(0.1)[$ = *[0, 9[*, with score *1*. Now the new score of A is *(55+1)/11 ≈ 5.09*, while the new score for B is: *(10+9+9+8+...+2+1)/11 = 64/11 ≈ 5.82*. Again we have a counterexample.

Note: $\widehat{Q_{101}}(p) = \widehat{P_{101}}(p)$ for *p* = *0.01, 0.02, …, 0.99* as $\widehat{Q_{101}}(p)$ is continuous for *p* = *0.01, 0.02, …, 0.99*. Indeed, $\widehat{Q_{101}}(x)$ is discontinuous in points x of the form m/101, with m a natural number between 1 and 100. Now m/101 is a percentile if it has the form q/100 with q a natural number between 0 and 100. In that case, one has 101*q = 100*m. Clearly, 100*m is a multiple of 100. As 101 is a prime number, 100 and 101 have no divisors in common. So 101*q = 100*m only if q is a multiple of 100. As q < 100 this is not possible. This shows that $\widehat{Q_{101}}(p)$ is continuous for *p* = *0.01, 0.02, …, 0.99* and hence, in these points : $\widehat{Q_{101}}(p) = \widehat{P_{101}}(p)$.